\documentclass[11pt]{article}

\usepackage[a4paper,margin=1in]{geometry}

\usepackage[T1]{fontenc}

\usepackage[utf8]{inputenc}

\usepackage{lmodern}

\usepackage{amsmath,amssymb}

\usepackage{graphicx}

\usepackage{booktabs}

\usepackage{microtype}

\usepackage{float}

\usepackage{enumitem}

\usepackage{xurl}

\usepackage[
    font=small,
    labelfont=bf,
    justification=justified,
    singlelinecheck=false
]{caption}

\usepackage{subcaption}

\usepackage[hidelinks]{hyperref}

\usepackage[numbers,sort&compress]{natbib}

\title{Boundary crossings as a minimal mechanism for global connectivity in structured landscapes}

\author{Fabio Nelli\\
Independent Researcher, Rome, Italy}

\date{}

\begin{document}

\maketitle

\begin{abstract}

Global connectivity in complex systems often emerges from local interactions, yet it remains unclear whether sophisticated agent strategies are required for such reorganisation. Here we identify a minimal mechanism by which heterogeneous mobility can generate system-wide connectivity when agents move within a structured multi-attractor landscape.

In an agent-based model, a small fraction of higher-mobility agents explores the environment while the majority remains locally constrained. Configurational transitions are strongly localised at inter-attractor boundaries, where local crossing events accumulate over time to produce an expanding transition network and a percolation-like shift from fragmented to globally connected regimes.

The core phenomenon persists under pure random-walk dynamics, showing that neither optimisation nor goal-directed behaviour is necessary. Interface-sensitive exploration nevertheless increases the transition rate by approximately four- to fivefold per exploratory movement relative to the random-walk baseline, thereby accelerating network integration. A novelty term produces only a small and non-uniform additional rate increase and no consistent improvement in unique-edge production. By contrast, when landscape structure is removed, transition-network connectivity arises through unconstrained spatial mixing, eliminating the boundary-mediated interpretation of the process.

These results identify a minimal mechanism through which simple stochastic movement and spatial constraints generate large-scale connectivity. Interfaces between basins emerge as critical mediators linking local dynamics to global organisation, with potential implications for systems in which movement occurs across structured environments.

\end{abstract}

\section{Introduction}

Understanding how local dynamics generate large-scale structural change remains a central problem in the study of complex systems. Across ecological, physical, social, and computational systems, global organisation can emerge from decentralised interactions among heterogeneous components without centralised control or explicit coordination \cite{holland1998,kauffman1993,camazine2001,prokopenko2009,epstein1999}. Identifying minimal mechanisms capable of producing such emergent reconfiguration remains a key challenge.

Agent-based modelling provides a natural framework for addressing this problem by allowing simple behavioural rules to be related systematically to collective dynamics \cite{bonabeau2002,wilensky2015}. In many contexts, system-level change is associated with the interplay between exploration and constraint. Exploration is often represented through strategic behaviour, adaptive learning, or cognitively guided search over complex landscapes \cite{march1991,levinthal1997,gavetti2000}. This raises a more basic question: can global reconfiguration emerge without agents optimising, learning, or explicitly seeking novelty?

Many complex systems can be represented as landscapes containing multiple basins of attraction separated by boundaries \cite{waddington1957,huang2012}. Movement within a basin tends to preserve local configurations, whereas crossing a boundary enables transitions between distinct states \cite{scheffer2009,huang2012}. Interfaces between basins may therefore play a disproportionate role in converting local movement into system-level change.

The ingredients of this problem have direct empirical counterparts. Large-scale human-mobility data reveal systematic heterogeneity between individuals whose trajectories concentrate on recurrent locations and individuals who explore a broader set of destinations \cite{song2010,pappalardo2015,alessandretti2018}. Mobility between cognitive domains is likewise observable in scientific careers: researchers change topics and disciplines, and these individual changes can be aggregated into networks or flows between fields \cite{jia2017,dedomenico2016,aleta2019,eykens2023}. At the network level, empirical communication data show that relatively sparse inter-community ties can be disproportionately important for maintaining global connectivity \cite{onnela2007}. Most recently, a rank-filtered analysis of 48 months of mobile-phone data from eight US cities identified a reproducible percolation threshold for the strongly connected backbone of urban origin--destination networks \cite{zhang2026}.

These studies do not validate the particular dynamics or thresholds examined here. They instead establish empirical counterparts for its principal ingredients: heterogeneous mobility, recurrently occupied domains, boundary-crossing transitions, and the system-level importance of connections between otherwise differentiated regions. The unresolved generative question is whether these independently observed ingredients are sufficient, in combination, to produce global connectivity.

Here we investigate whether heterogeneous mobility interacting with such spatial structure is sufficient to generate global connectivity. We consider agents moving in a continuous multi-attractor landscape, with a small fraction exhibiting higher-mobility exploratory dynamics while the remainder remain locally constrained. In the minimal case, exploratory agents follow simple random-walk dynamics and neither optimise their trajectories nor explicitly seek new states \cite{codling2008}.

To isolate the mechanisms involved, we compare this baseline with progressively enriched dynamics incorporating interface sensitivity and novelty-biased exploration \cite{march1991,levinthal1997,gavetti2000}, together with a flat-landscape control in which spatial structure is removed. Global organisation is characterised through the transition network generated by observed movements between attractor regions \cite{newman2018,barabasi2016,boccaletti2006}.

We find that increasing the exploratory fraction progressively connects previously separated regions of the landscape. Transition events are strongly localised near inter-attractor boundaries \cite{scheffer2009,huang2012}, and their accumulation produces an expanding transition network that shifts from fragmented to globally connected regimes. Boundary localisation is shared across exploratory mechanisms, but interface-sensitive selection increases the transition rate by approximately four- to fivefold per exploratory movement relative to the random-walk baseline. The same qualitative mechanism persists under pure random-walk exploration, whereas removing landscape structure eliminates its boundary-mediated interpretation.

These results identify a minimal generative mechanism for global connectivity \cite{epstein1999}: simple stochastic movement, when heterogeneous and constrained by a structured environment, can transform local boundary crossings into system-wide connectivity. The results therefore shift attention from agent-level sophistication to the interaction between mobility and environmental structure, with interfaces acting as mediators between local dynamics and global organisation.

\section{Model}

\subsection{System overview}

We consider an agent-based model defined on a continuous two-dimensional space

\[\mathcal{X} \subset \mathbb{R}^2,\]

in which a population of \(N\) agents evolves over discrete time steps

\[t = 0,1,\ldots,T.\]

Each agent occupies a position

\[x_i(t) \in \mathcal{X}\]

and is associated with a discrete state corresponding to the attractor region to which it belongs. The environment contains \(K\) attractors,

\[\mathcal{A}=\{A_1,A_2,\ldots,A_K\},\]

with centres \(c_k \in \mathbb{R}^2\) and associated regions of attraction. This construction defines a structured multi-basin landscape in which neighbouring basins are separated by interfaces. Such landscapes provide a minimal representation of systems with multiple stable or metastable configurations \cite{kauffman1993,scheffer2009,waddington1957,huang2012}.

Agents are divided into two types. Each agent is assigned independently to the exploratory type with probability

\[p \in [0,1]\]

and otherwise to the embedded type. Consequently, the realised number of exploratory agents is stochastic,

\[N_e\sim\operatorname{Binomial}(N,p),
\qquad
\mathbb{E}[N_e]=Np.\]

The distinction between the two populations is implemented through their movement-update rules. Agents do not communicate, receive payoffs, or perform explicit goal-directed planning.

Initial positions are sampled uniformly from attractor-centred bounding squares. The initial state is then assigned independently by the nearest-centre rule. The complete initialization procedure, landscape geometry, and parameter specifications are provided in the Supplementary Methods.

\subsection{Agent dynamics}

In the baseline regime, the two agent types differ exclusively in the scale and constraint of their stochastic motion. Embedded agents follow locally constrained dynamics biased toward the centre of their current attractor. Their position evolves according to

\[\mathbf{x}_i(t+1)
=
\mathbf{x}_i(t)
+
\gamma_k\bigl(\mathbf{c}_k-\mathbf{x}_i(t)\bigr)
+
\boldsymbol{\epsilon}_i(t),\]

where \(\mathbf{c}_k\) denotes the centre of the attractor to which the agent currently belongs, \(\gamma_k\) controls the attraction strength, and

\[\boldsymbol{\epsilon}_i(t)
=
\bigl(\epsilon_{i,x}(t),\epsilon_{i,y}(t)\bigr),
\qquad
\epsilon_{i,\alpha}(t)
\overset{\mathrm{iid}}{\sim}
\mathcal{U}(-\delta_b,\delta_b),
\quad
\alpha\in\{x,y\}\]

is a small stochastic perturbation. This dynamics produces trajectories that remain confined within, or fluctuate around, an attractor basin.

After every proposed displacement, positions are projected component-wise onto the rectangular domain \(\mathcal{X}=[0,W]\times[0,H]\),

\[\Pi_{\mathcal{X}}(x,y)
=
\bigl(\min\{\max\{x,0\},W\},
      \min\{\max\{y,0\},H\}\bigr).\]

Thus, the implementation uses clamped boundaries rather than periodic or specularly reflecting boundaries. The movement equations in this section are understood to be followed by this projection.

In the baseline exploratory regime, exploratory agents instead follow a higher-mobility random walk,

\[\mathbf{x}_i(t+1)
=
\mathbf{x}_i(t)
+
\boldsymbol{\eta}_i(t),\]

with

\[\boldsymbol{\eta}_i(t)
=
\bigl(\eta_{i,x}(t),\eta_{i,y}(t)\bigr),
\qquad
\eta_{i,\alpha}(t)
\overset{\mathrm{iid}}{\sim}
\mathcal{U}(-\delta_e,\delta_e),
\quad
\alpha\in\{x,y\},\]

and

\[\delta_e>\delta_b.\]

Their trajectories are therefore more diffusive and can cross boundaries between attractor regions. This dynamics is consistent with standard random-walk descriptions of undirected movement and exploration \cite{codling2008}.

In the enriched exploratory regimes, agents generate \(M\) independently sampled candidate displacements using the same exploratory increment distribution and select the candidate with the largest local score. Let \(\mathbf{x}_{im}^{\mathrm{cand}}\) be candidate \(m\), let \(s(\mathbf{x})\) denote its nearest-attractor state, and let \(\mathcal{H}_i(t)\) be the set of attractor states previously visited by agent \(i\). The implemented score is

\[S_{im}(t)
=
\beta_I\,
\mathbf{1}\!\left[d_B\!\left(\mathbf{x}_{im}^{\mathrm{cand}}\right)
\leq \epsilon_I\right]
+
\beta_N\,
\mathbf{1}\!\left[s\!\left(\mathbf{x}_{im}^{\mathrm{cand}}\right)
\notin\mathcal{H}_i(t)\right]
+\xi_{im},\]

where \(d_B\) is the Voronoi-boundary distance defined below and \(\xi_{im}\) is an infinitesimal random term used only to break exact ties. The interface contribution is therefore a binary bonus for candidates inside the prescribed interface zone, not a continuous function of distance. The interface-only condition sets \(\beta_N=0\); the full condition uses both terms. Candidate generation contains no attractor-directed drift.

\subsection{Experimental mechanisms}

To isolate the contribution of different exploratory mechanisms, we compare three experimental regimes of increasing structure.

\begin{enumerate}[label=\textbf{\arabic*.},leftmargin=1.2cm]

\item \textbf{Baseline (random walk).}

Exploratory agents follow the higher-mobility random-walk dynamics defined above, with no interface sensitivity, novelty bias, or candidate selection \cite{codling2008}.

\item \textbf{Interface-only.}

Exploratory agents generate multiple unbiased stochastic candidate displacements and select among them using only the binary interface-zone bonus. This increases sampling of inter-attractor interfaces while preserving stochastic exploration.

\item \textbf{Full model (interface + novelty).}

Exploratory agents use the same unbiased candidate-generation procedure, but candidates are scored using both the binary interface-zone bonus and a novelty bonus for attractor regions not previously visited. This introduces a minimal history-dependent component without attractor-directed drift, explicit optimisation, or goal-directed planning \cite{march1991,levinthal1997,gavetti2000}.

\end{enumerate}

This decomposition distinguishes effects arising from mobility heterogeneity alone from those introduced by interface-sensitive candidate selection and the additional novelty term. The complete candidate-generation and scoring procedure is provided in the Supplementary Methods.

\subsection{Landscape structure}

The landscape is considered under two configurations. In the structured configuration, multiple attractors generate spatially distinct basins separated by dynamically meaningful interfaces \cite{scheffer2009,waddington1957,huang2012}. Agent motion is therefore constrained by spatial gradients associated with the attractor structure.

As a null spatial control, we consider a flat-landscape configuration in which attractor-directed drift, interface-sensitive candidate selection, and novelty-dependent selection are disabled. Embedded and exploratory agents then perform unbiased stochastic motion with their respective displacement scales, while the fixed nearest-attractor partition is retained solely for state assignment and transition detection. Agent transitions can therefore still be recorded, but the regions no longer correspond to dynamically separated basins.

Comparison between the structured and flat configurations allows us to distinguish boundary-mediated connectivity from connectivity arising through unconstrained spatial mixing. The same state-assignment and transition-detection procedures are applied in both configurations.

\subsection{Transition graph construction}

To characterize system-level connectivity, we construct a directed transition graph

\[G=(V,E),\]

whose nodes correspond to attractor regions and whose edges are generated directly from observed agent trajectories \cite{newman2018,barabasi2016,boccaletti2006}.

At each time step, every agent is assigned to its nearest attractor centre. A directed edge

\[i \rightarrow j\]

is included whenever at least one transition from attractor \(A_i\) to attractor \(A_j\) is observed during the simulation. Edge weights record the cumulative number of transitions between each ordered pair of attractors.

The same graph-construction procedure is applied across all experimental conditions, including the flat-landscape control. The transition graph therefore provides a common representation through which the accumulation of local state transitions can be related to system-level connectivity.

The complete mathematical definition of transition counts and edge construction is provided in the Supplementary Methods.

\subsection{Observables}

We characterize the system using observables defined at three levels: spatial organisation, transition events, and network connectivity.

\subsubsection*{Spatial observables}

The attractor centres induce a nearest-centre Voronoi partition. Let

\[s(\mathbf{x})
=
\operatorname*{arg\,min}_{k\in\{1,\ldots,K\}}
\lVert\mathbf{x}-\mathbf{c}_k\rVert,\]

with ties resolved deterministically by attractor order. For a position \(\mathbf{x}\) assigned to basin \(k=s(\mathbf{x})\), the implementation measures the exact perpendicular distance to the nearest Voronoi boundary:

\[d_B(\mathbf{x})
=
\min_{j\neq k}
\frac{
\left|
\lVert\mathbf{x}-\mathbf{c}_j\rVert^2
-
\lVert\mathbf{x}-\mathbf{c}_k\rVert^2
\right|
}{2\lVert\mathbf{c}_j-\mathbf{c}_k\rVert}.\]

The denominator converts the difference in squared distances into perpendicular distance from the equal-distance bisector. Small values identify positions close to an inter-basin interface, whereas larger values identify positions deeper within the current basin. This Voronoi metric, rather than distance from an attractor's nominal initialization radius, is used throughout the spatial and event-level analyses.

Spatial organisation is summarized separately for embedded and exploratory agents by

\[\overline{d}_{B,g}(t)
=
\frac{1}{N_g}
\sum_{i:\,g_i=g}d_B\!\left(\mathbf{x}_i(t)\right),
\qquad
g\in\{\mathrm{embedded},\mathrm{exploratory}\},\]

and by interface occupation. An observation lies in the interface zone when

\[d_B\!\left(\mathbf{x}_i(t)\right)\leq\epsilon_I.\]

The reported interface ratio is the mean, over agents of a given type, of the fraction of their observed moves ending in this zone.

\subsubsection*{Event-level observables}

At each discrete time step, a transition is operationally defined by a change in nearest-attractor assignment between consecutive positions,

\[Y_i(t)
=
\mathbf{1}\!\left[
s\!\left(\mathbf{x}_i(t+1)\right)
\neq
s\!\left(\mathbf{x}_i(t)\right)
\right].\]

The procedure records endpoint state changes and does not reconstruct every possible geometrical intersection along the intervening displacement segment. To determine where these changes occur, we measure

\[P\!\left(\mathrm{transition}\mid d_{B,\mathrm{pre}}\right)
=
\mathbb{P}\!\left[
Y_i(t)=1
\mid
d_B\!\left(\mathbf{x}_i(t)\right)=d_{B,\mathrm{pre}}
\right].\]

For each simulation run and distance bin, this probability is estimated as the number of pre-move observations followed by a transition divided by the total number of pre-move observations in that bin. The plotted estimate is the unweighted mean of these run-level probabilities over runs containing at least one observation in the bin,

\[\widehat P_b
=
\frac{1}{R_b}
\sum_{r\in\mathcal{R}_b}
\frac{N^{\mathrm{trans}}_{rb}}{N^{\mathrm{obs}}_{rb}},\]

where \(\mathcal{R}_b\) is the set of contributing runs and \(R_b=|\mathcal{R}_b|\). Shaded bands are normal-approximation intervals across run-level estimates, \(\widehat P_b\pm1.96s_b/\sqrt{R_b}\). The dedicated analysis pools \(p=0.01,0.02,0.03\), with 20 independent runs per value. For descriptive event distributions, the event distance is recorded as the smaller of the pre- and post-move boundary distances. The event-density histogram pools individual events across these conditions and is therefore weighted by the number of events produced in each condition. The conditional probability is the stronger localisation diagnostic because it controls for the number of observations available at each distance.

To separate transition frequency from topological output, we additionally define three run-level exploratory measures. Let \(M_e\) be the total number of movements made by exploratory agents, \(T_e\) the number of exploratory inter-basin transitions, and \(U_e\) the number of distinct directed edges generated by those transitions. The exploratory transition rate is

\[\lambda_e=\frac{T_e}{M_e},\]

the production of distinct directed edges per exploratory movement is

\[\gamma_e=\frac{U_e}{M_e},\]

and the unique-edge yield is

\[\eta_e=\frac{U_e}{T_e}.\]

The final quantity is equivalently the fraction of exploratory transitions that constitute the first observed occurrence of a directed edge within a run. These measures satisfy \(\gamma_e=\lambda_e\eta_e\) at the run level. Ratios with a zero denominator are treated as undefined rather than assigned a value of zero. Because the main experiment uses \(K=3\), \(U_e\leq K(K-1)=6\); consequently, \(\gamma_e\) and \(\eta_e\) necessarily decline once the small directed graph approaches saturation and should be interpreted primarily through within-\(p\) comparisons between mechanisms.

\subsubsection*{Network observables}

The transition graph is characterized using standard network measures \cite{newman2018,barabasi2016,boccaletti2006}. Graph density is defined as

\[D=
\frac{|E|}
{|V|(|V|-1)},\]

and the mean total degree as

\[\langle k_{\mathrm{tot}}\rangle
=
\frac{2|E|}
{|V|}.\]

Large-scale connectivity is quantified by the relative sizes of the largest weakly and strongly connected components,

\[C_w
=
\frac{|V_{\mathrm{largest\ weak}}|}
{|V|},
\qquad
C_s
=
\frac{|V_{\mathrm{largest\ strong}}|}
{|V|}.\]

The growth of these components provides a measure of the progressive integration of attractor regions. In particular, \(C_w\) is used to characterize the emergence of system-wide connectivity, which can be compared with connectivity growth in percolation processes \cite{stauffer2003}.

\section{Experimental Design and Methodology}

\subsection{Parameter space and simulation protocol}

We investigate the system dynamics by varying the exploratory-agent fraction over

\[p \in [0,0.05],\]

in increments of \(0.005\). For each value of \(p\), we perform \(R=20\) independent simulations using different random seeds to account for stochastic variability.

The main experiments use \(N=200\) agents, \(K=3\) attractors, and a simulation duration of \(T=300\) time steps. This deliberately minimal state space provides the simplest configuration in which cross-basin network integration can be measured, but connectivity measures can saturate rapidly when only three attractor regions are present. Robustness analyses therefore vary the number of attractors, landscape scaling, simulation duration, and population size, including substantially larger transition networks. The principal model parameters are reported in Table~\ref{tab:parameters}, and the additional geometrical configurations and system sizes are described in the Supplementary Methods.

\begin{table}[H]

\centering

\caption{Principal parameters of the main simulation experiments. The nominal attractor radius is used for initialization and does not define the Voronoi basin boundary.}

\label{tab:parameters}

\begin{tabular}{lll}
\toprule
Parameter & Symbol & Value \\
\midrule
Domain width and height & \(W,H\) & \(50,50\) \\
Number of agents & \(N\) & \(200\) \\
Number of attractors & \(K\) & \(3\) \\
Exploratory probability & \(p\) & \(0\)--\(0.05\), step \(0.005\) \\
Simulation duration & \(T\) & \(300\) steps \\
Independent realizations & \(R\) & \(20\) \\
Attraction strength & \(\gamma_k\) & \(0.30\) \\
Embedded displacement scale & \(\delta_b\) & \(0.5\) \\
Exploratory displacement scale & \(\delta_e\) & \(2.0\) \\
Nominal initialization radius & \(r_k\) & \(8.0\) \\
Interface-zone threshold & \(\epsilon_I\) & \(3.0\) \\
Candidates per exploratory update & \(M\) & \(5\) \\
Interface bonus & \(\beta_I\) & \(0.5\) \\
Novelty bonus & \(\beta_N\) & \(1.0\) \\
\bottomrule\end{tabular}

\end{table}

For an observable \(X\), results at each value of \(p\) are averaged across independent realizations,

\[\langle X(p)\rangle
=
\frac{1}{R}
\sum_{r=1}^{R}
X_r(p),\]

where \(X_r(p)\) denotes the value measured in realization \(r\). Independent replication and controlled random seeds are used throughout the experimental pipeline, following standard practice in stochastic agent-based modelling \cite{bonabeau2002,wilensky2015}.

\subsection{Mechanism isolation framework}

The experimental design follows the three exploratory regimes defined in Section~2.3: baseline random walk, interface-only candidate selection, and the full interface-plus-novelty model. These conditions progressively introduce interface-sensitive candidate selection and memory-dependent novelty, allowing their contributions to be distinguished from mobility heterogeneity alone \cite{codling2008,march1991,levinthal1997,gavetti2000}.

In addition, the flat-landscape control defined in Section~2.4 removes attractor-induced spatial structure while preserving the transition-detection procedure. Together, these comparisons isolate the respective contributions of heterogeneous mobility, landscape structure, and exploratory selection mechanisms.

\subsection{Metrics and statistical analysis}

For each experimental condition, we compute the spatial, event-level, and network observables defined in Section~2.6. Spatial differentiation is assessed separately for embedded and exploratory agents using interface occupation and Voronoi-boundary distance. Transition frequency and topological output are separated using \(\lambda_e\), \(\gamma_e\), and \(\eta_e\), each calculated within an individual run before aggregation. Network organisation is characterized by the number of observed edges, graph density, mean total degree, and the relative sizes of the largest weakly and strongly connected components \cite{newman2018,barabasi2016,boccaletti2006}.

All observables are computed independently for each simulation run and subsequently averaged across realizations according to the statistical protocol described in Section~3.1. Conditions use paired seeds for each \((p,r)\) combination, matching initialization, exploratory-type assignments, and initial positions before their movement histories diverge. For the three exploratory rate and yield measures, paired condition contrasts are summarized by the mean run-level difference and a percentile 95\% confidence interval obtained from 10,000 bootstrap resamples of the valid run pairs. Replicates with no exploratory movements do not contribute to movement-normalized ratios, and replicates with no exploratory transitions do not contribute to \(\eta_e\). These exclusions affect principally the two smallest positive values of \(p\). The pointwise intervals are descriptive and are not adjusted for multiple comparisons.

\subsection{Detection of regime transitions}

We characterize changes in system behaviour from the joint dependence of spatial and network observables on the exploratory fraction \(p\), without assuming a sharp phase transition. Particular attention is given to the growth of the largest weakly connected component \(C_w\), which measures the progressive integration of attractor regions.

We refer to the resulting connectivity shift as percolation-like because system-wide connectivity emerges through the accumulation of local transitions. Unlike classical percolation on a predefined static network, however, the transition graph here is generated endogenously by agent dynamics \cite{newman2018,barabasi2016,boccaletti2006,stauffer2003}.

\subsection{Reproducibility}

All simulations are implemented in Python using Mesa for agent-based execution and NetworkX for transition-network analysis. Random seeds are controlled and recorded, and model parameters and experimental configurations are specified externally. Data generation, aggregation, and figure production are performed through a common computational pipeline to ensure consistency across experimental conditions.

\section{Results}

\subsection{Spatial segregation in structured landscapes}

We first examine how heterogeneous mobility shapes spatial organisation within the structured landscape. In the baseline configuration, exploratory agents follow random-walk dynamics, whereas embedded agents remain locally constrained around attractor centres.

\begin{figure}[H]

    \centering

    \includegraphics[width=\textwidth]{\detokenize{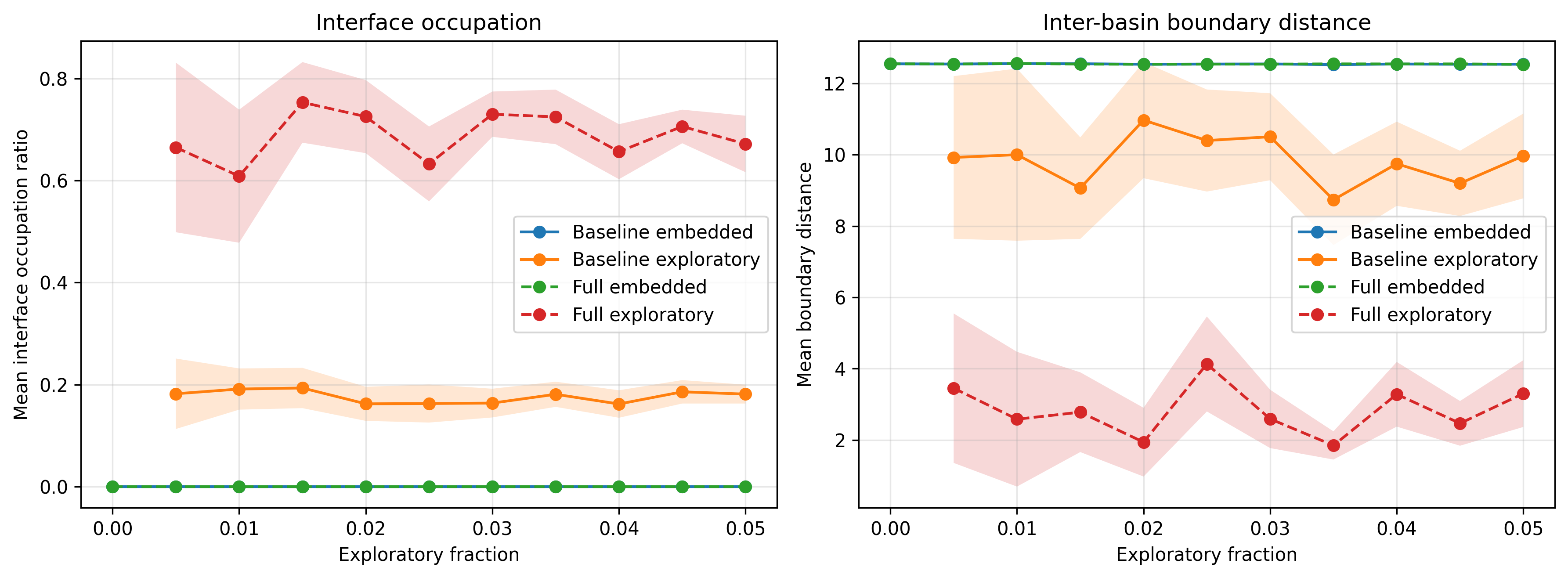}}

    \caption{Spatial differentiation between embedded and exploratory agents. Interface occupation and mean Voronoi-boundary distance as functions of the exploratory fraction \(p\), shown for the baseline and full-model configurations. Shaded regions are approximate 95\% confidence intervals across independent runs.}

    \label{fig:spatial}

\end{figure}

As shown in Figure~\ref{fig:spatial}, the two populations occupy distinct regions of the landscape. Embedded agents remain predominantly within attractor basins and maintain larger distances from inter-attractor boundaries, whereas exploratory agents exhibit lower mean boundary distances and higher interface occupation.

This differentiation is already present under baseline random-walk exploration, showing that mobility heterogeneity alone is sufficient to generate spatial differentiation in a structured landscape. Interface-sensitive exploration strengthens the effect but does not qualitatively change it. This is consistent with the broader observation that heterogeneous movement can generate spatial organisation without explicit coordination \cite{codling2008,camazine2001}.

\subsection{Transition events localise at boundaries}

Spatial differentiation alone does not establish whether interfaces mediate configurational change. We therefore examine where transitions between attractor states occur relative to inter-attractor boundaries.

\begin{figure}[H]

    \centering

    \includegraphics[width=\textwidth]{\detokenize{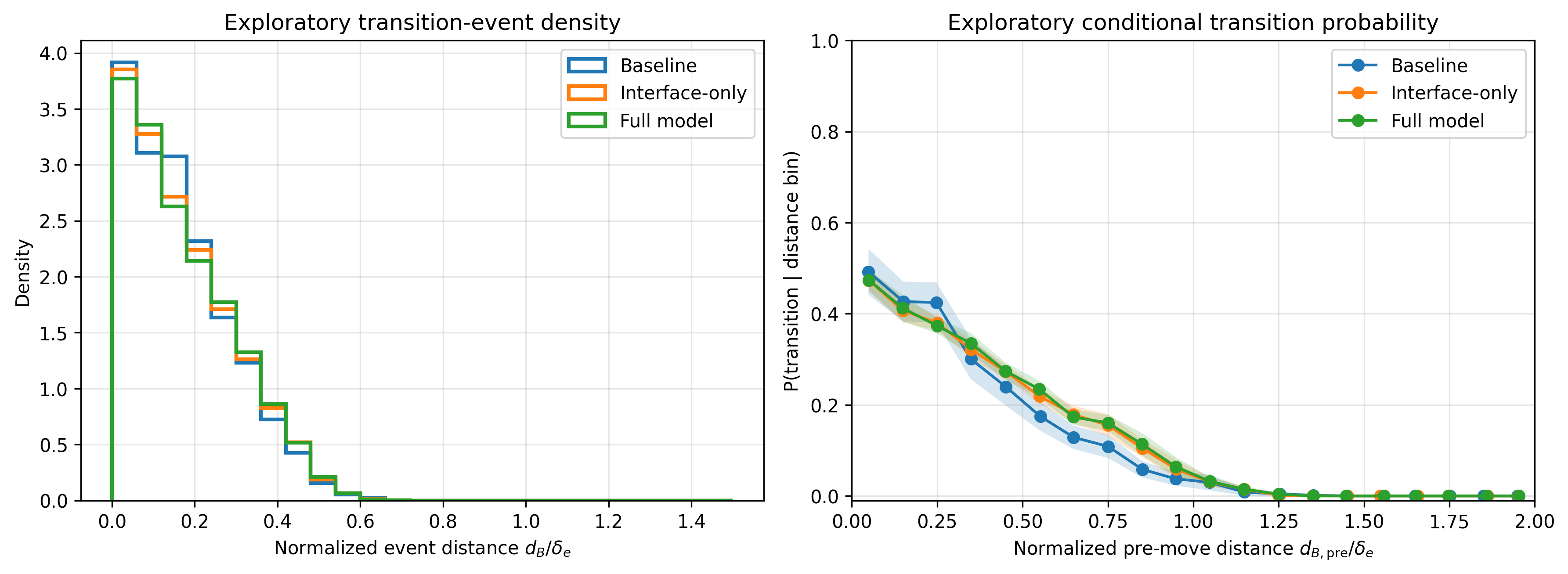}}

    \caption{Boundary localisation of exploratory transition events across the baseline random-walk, interface-only, and full interface-plus-novelty conditions. Dedicated analyses pool \(p=0.01,0.02,0.03\), with 20 runs per value and condition. The left panel shows event-weighted distributions of exact Voronoi-boundary distance, normalized by the exploratory movement scale \(\delta_e\). The right panel shows mean run-level conditional transition probabilities as functions of normalized pre-move boundary distance, focusing on the transition-relevant interval \(0\leq d_{B,\mathrm{pre}}/\delta_e\leq2\). Shaded bands are 95\% normal-approximation intervals across contributing run-level estimates.}

    \label{fig:events}

\end{figure}

As shown in Figure~\ref{fig:events}, the event-distance distributions are nearly coincident across the three exploratory mechanisms. Their mass is concentrated close to the nearest inter-basin boundary and becomes negligible beyond approximately \(0.6\delta_e\). Boundary localisation is therefore not created specifically by interface-sensitive selection: it is a geometric property of changing nearest-attractor state.

The conditional transition probability provides the stronger localisation test because it controls for the number of available observations at each distance. Across all three mechanisms, this probability is approximately one-half in the boundary-adjacent bin, decreases monotonically with normalized pre-move distance, and approaches zero by approximately \(1.2\delta_e\). The curves show only modest separation and their uncertainty bands overlap over much of the range.

Interfaces therefore constitute dynamical loci of configurational switching rather than merely regions of exploratory occupation. The geometry determines where a crossing can occur, whereas the exploratory rule determines how frequently agents are exposed to those crossing-enabling regions. Local fluctuations occurring near basin boundaries are preferentially converted into transitions between attractor states, consistent with theoretical descriptions of switching between metastable configurations \cite{kauffman1993,scheffer2009,huang2012}.

\subsection{From localised switching to transition-network expansion}

We next examine how local transition events accumulate at the system level through the transition graph defined in Section~2.5.

\begin{figure}[H]

    \centering

    \includegraphics[width=\textwidth]{\detokenize{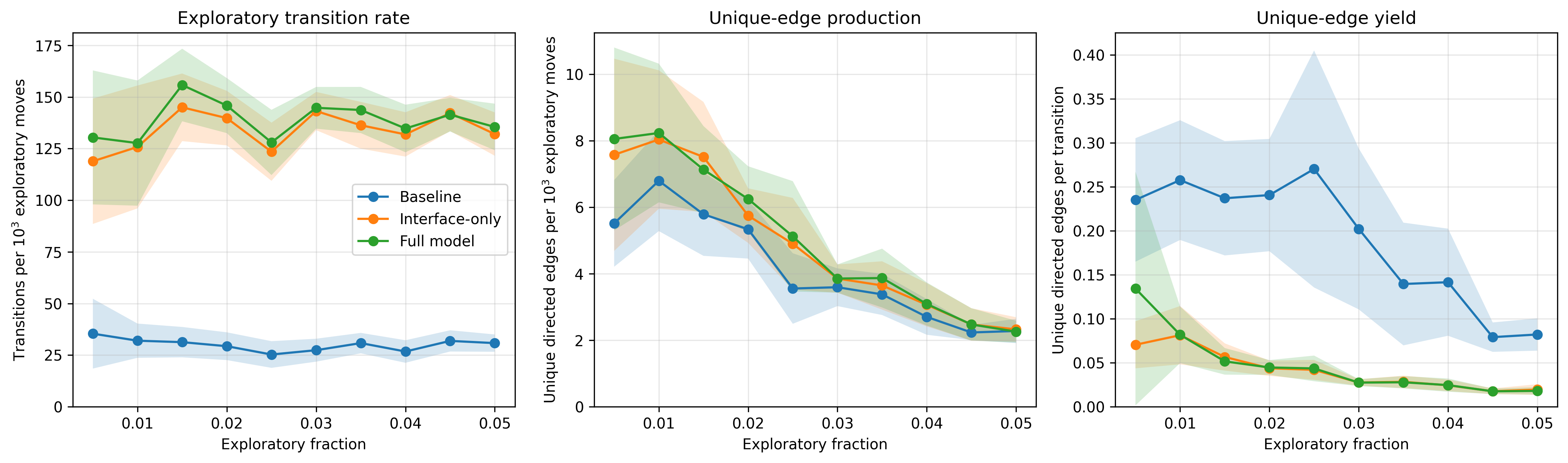}}

\caption{Transition frequency and topological output across exploratory mechanisms. The left panel shows exploratory transitions per \(10^3\) exploratory movements. The centre panel shows distinct directed edges produced per \(10^3\) exploratory movements. The right panel shows the fraction of exploratory transitions yielding a previously unobserved directed edge within the run, \(\eta_e=U_e/T_e\). Metrics are calculated within each run before aggregation; runs with a zero denominator do not contribute to the corresponding ratio. Shaded regions are 95\% normal-approximation intervals across contributing runs.}

    \label{fig:network_expansion}

\end{figure}

As shown in Figure~\ref{fig:network_expansion}, interface-sensitive candidate selection produces a large and persistent increase in crossing frequency after controlling for the number of exploratory movements. Relative to the baseline random walk, the interface-only condition generates an additional 83.5--115.8 transitions per \(10^3\) exploratory movements across the tested positive values of \(p\). The paired 95\% bootstrap interval excludes zero at every value from \(p=0.005\) to \(0.05\). Given baseline rates of approximately 25--35 transitions per \(10^3\) movements, this corresponds to an approximately four- to fivefold increase.

The corresponding advantage in distinct-edge production is concentrated in the pre-saturation regime. The interface-only minus baseline contrast in \(\gamma_e\) has a paired interval excluding zero at \(p=0.015\), \(0.025\), and \(0.040\), but not consistently elsewhere. Because only six directed edges are possible for \(K=3\), the numerator \(U_e\) saturates while movements and repeated transitions continue to accumulate. Consistent with this ceiling effect, \(\eta_e\) is lower under interface-only than under the baseline at every positive \(p\), with paired intervals excluding zero throughout. This does not imply that the baseline constructs a more connected graph; rather, interface seeking generates many more crossings, an increasing fraction of which repeat edges that have already been observed.

The novelty term produces a smaller secondary effect. The full model exceeds interface-only in transition rate at six of the ten positive values of \(p\) under the pointwise paired bootstrap intervals, with supported differences of approximately 3.5--11.5 transitions per \(10^3\) movements. It improves distinct-edge production at only one value, \(p=0.020\), and produces no consistent change in unique-edge yield. These pointwise comparisons are descriptive and unadjusted for multiplicity; they support a modest, non-uniform rate effect rather than a separate topological mechanism.

Combined with the event-level localisation shown in Figure~\ref{fig:events}, these results identify the operative sequence: interface-sensitive selection increases exposure to crossing-enabling regions, this exposure raises the number of boundary crossings per unit movement, and repeated local events expand the transition network. The progression from local transitions to network-level structure is consistent with the emergence of connectivity through repeated local events in complex networks \cite{newman2018,barabasi2016,boccaletti2006}.

\subsection{Emergence of global connectivity}

The expansion of the transition network is accompanied by a change in global connectivity, which we quantify through the relative sizes of the largest weakly and strongly connected components.

\begin{figure}[H]

    \centering

    \includegraphics[width=\textwidth]{\detokenize{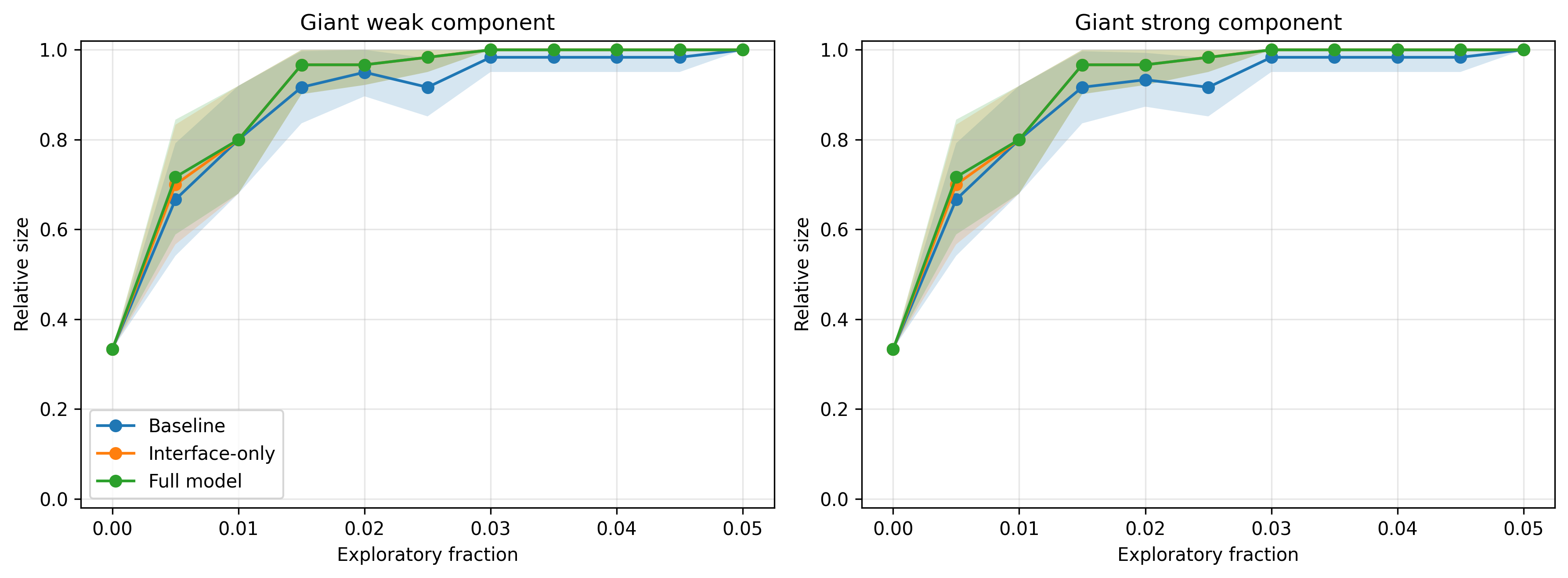}}

    \caption{Emergence of global connectivity. Relative sizes of the largest weakly and strongly connected components as functions of the exploratory fraction \(p\) across experimental conditions. Lines show across-run means and shaded regions show approximate 95\% confidence bands, clipped to the admissible interval \([0,1]\).}

    \label{fig:global_connectivity}

\end{figure}

As shown in Figure~\ref{fig:global_connectivity}, increasing the exploratory fraction shifts the transition network from a fragmented regime toward global integration. At low \(p\), transitions connect only subsets of the attractor landscape, whereas increasing \(p\) progressively merges these components and increases the fraction of attractor regions contained in the largest connected component.

This connectivity growth occurs already under baseline random-walk exploration. Interface-sensitive candidate selection modifies the approach to integration, while novelty is not required for the emergence of a system-spanning component. The emergence of a system-spanning component from accumulated local transitions is therefore consistent with a percolation-like connectivity process \cite{newman2018,barabasi2016,boccaletti2006,stauffer2003}.

Illustrative transition-network snapshots for \(K=10\) show how this integration develops structurally (Supplementary Figure~S1). At very low exploratory fractions the configurations remain disconnected, whereas at an intermediate exploratory fraction a system-spanning weakly connected component is present before the network becomes substantially denser at higher \(p\). Global connectivity and subsequent network densification are therefore distinguishable aspects of transition-network growth. The snapshots are individual predetermined-seed realisations and do not estimate a critical threshold; quantitative conclusions are based on the replicated aggregate analyses.

\subsection{Robustness across landscape size}

The main \(K=3\) experiment provides the minimal state space for observing network integration but also reaches topological saturation rapidly. We therefore examine whether the emergence of connectivity persists as the number of attractor regions increases.

\begin{figure}[H]

    \centering

    \includegraphics[width=\textwidth]{\detokenize{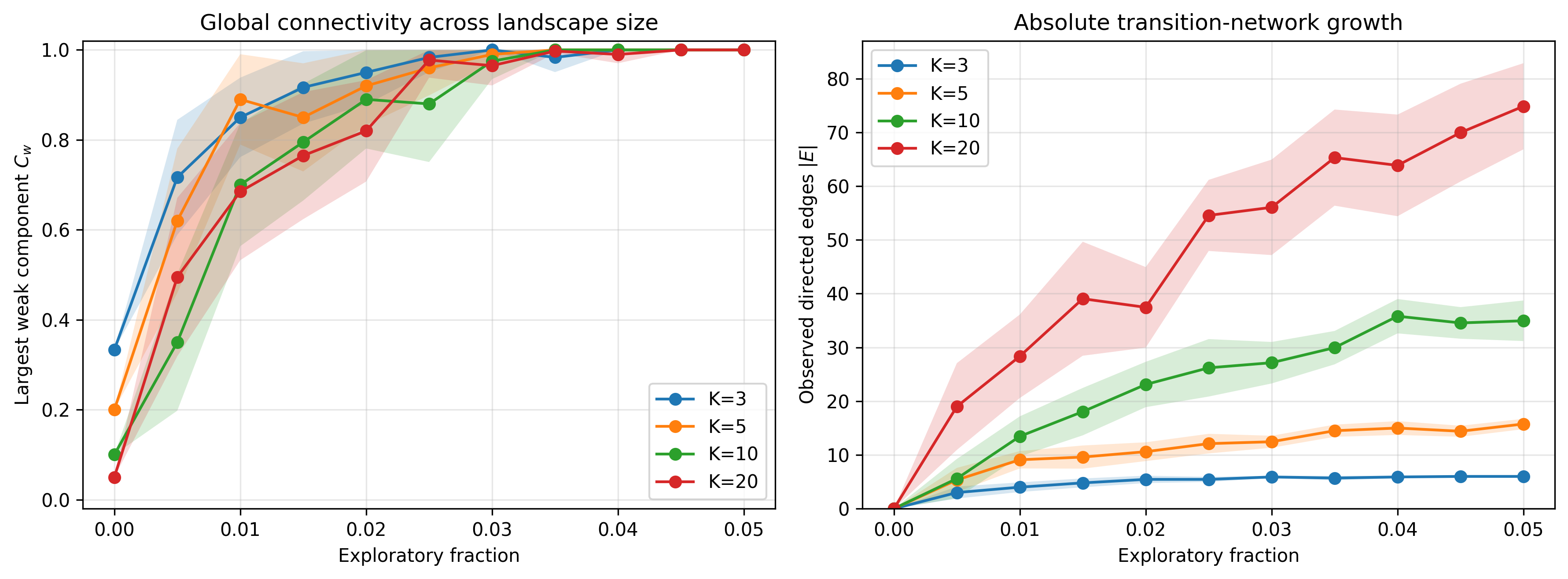}}

    \caption{Robustness across landscape size under native fixed-domain geometry. The left panel shows the relative size of the largest weakly connected component for \(K=3,5,10,20\). The right panel shows the corresponding absolute number of observed directed edges. At \(p=0\), the component size equals \(1/K\), as expected for isolated attractor nodes. Increasing \(K\) within a fixed domain changes basin number, spacing, occupancy, and movement scale relative to the geometry simultaneously; the constant-spacing robustness analysis in the Supplementary Material separates these effects. Shaded regions are approximate 95\% confidence intervals across independent runs.}

    \label{fig:k_robustness}

\end{figure}

Figure~\ref{fig:k_robustness} shows connectivity growth for every tested landscape size. The initial value satisfies \(C_w=1/K\), providing an internal check on the isolated-node configuration. Increasing \(p\) then expands the largest weak component toward system-wide connectivity for \(K=3,5,10,20\), while the absolute number of observed directed edges grows substantially with landscape size. Thus, the qualitative result is not restricted to the six-edge graph available at \(K=3\).

The curves are not expected to be ordered monotonically by \(K\) under this design. Because the domain and population are fixed, increasing \(K\) simultaneously changes attractor spacing, basin geometry, agents per basin, and the displacement scale relative to basin size. Figure~\ref{fig:k_robustness} should therefore be interpreted as native fixed-domain robustness rather than as an isolated causal effect of network size. The supplementary constant-spacing and finite-time analyses provide the complementary controls needed to distinguish landscape crowding from state-space size.

\subsection{Mechanism decomposition}

To distinguish the roles of mobility heterogeneity, interface sensitivity, and novelty, we compare the spatial observables across the progressively enriched exploratory regimes.

\begin{figure}[H]

    \centering

    \includegraphics[width=\textwidth]{\detokenize{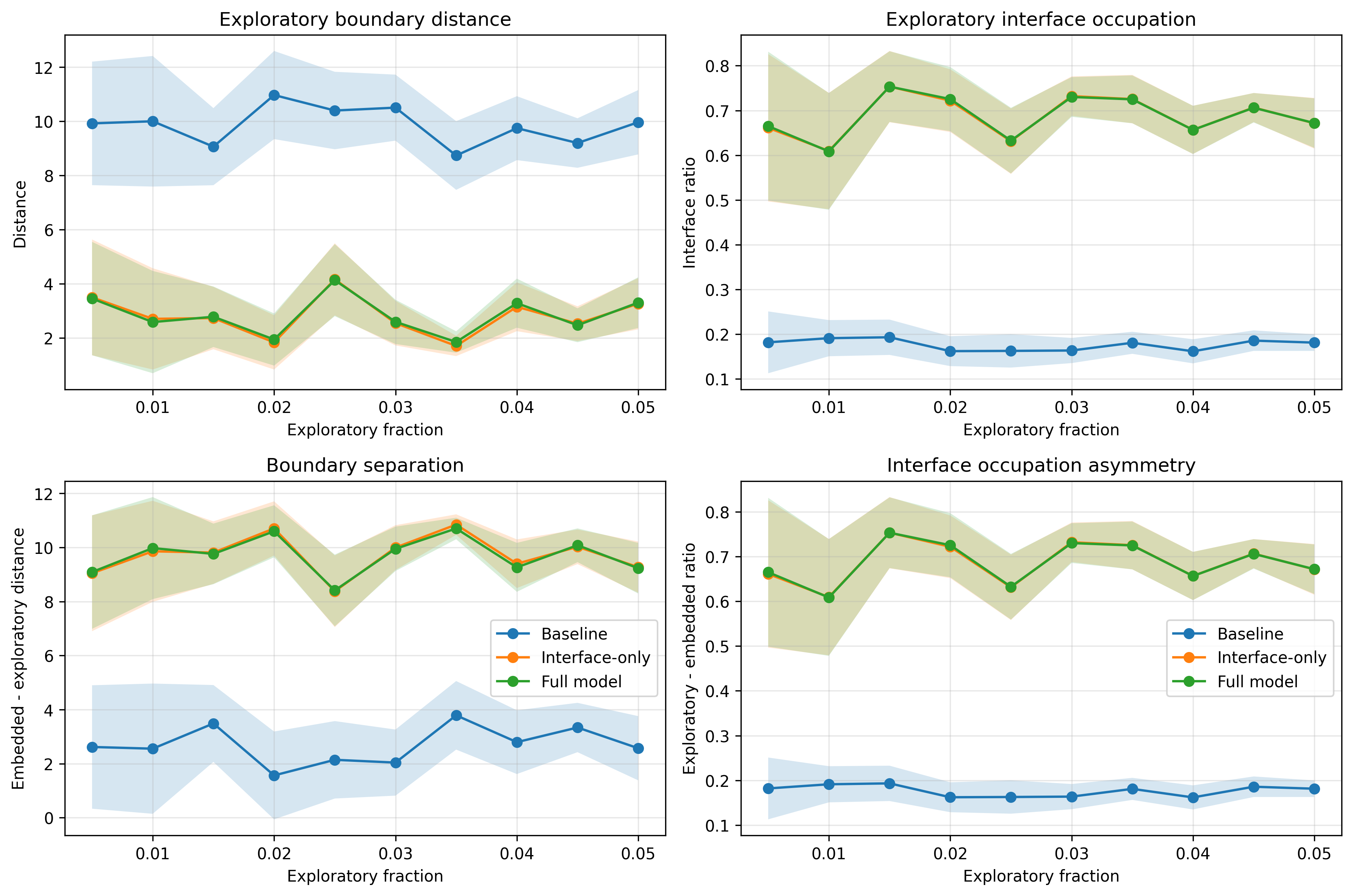}}

    \caption{Mechanism decomposition across the baseline, interface-only, and full-model conditions. Panels show exploratory boundary distance, exploratory interface occupation, boundary-distance separation between embedded and exploratory agents, and interface-occupation asymmetry. Shaded regions are approximate 95\% confidence intervals across independent runs.}

    \label{fig:mechanism_decomposition}

\end{figure}

Figure~\ref{fig:mechanism_decomposition} shows that mobility heterogeneity alone already produces a spatial distinction between embedded and exploratory agents. Introducing interface-sensitive candidate selection strongly increases exploratory occupation of inter-basin boundaries and decreases their mean distance from those boundaries. The interface-only and full-model curves are nearly coincident across the tested exploratory fractions; consequently, novelty produces no clearly distinguishable additional effect on these spatial observables.

The decomposition therefore separates the robust effects of mobility heterogeneity and interface sensitivity. Higher exploratory mobility is sufficient to generate cross-basin transitions and network growth, whereas interface sensitivity changes where exploratory motion is concentrated, making boundary regions the dominant loci of exploratory activity. The novelty term implements a minimal history-dependent bias toward previously unvisited regions and produces a small transition-rate increment at some values of \(p\) (Figure~\ref{fig:network_expansion}), but it does not generate a consistent improvement in distinct-edge production or a separate spatial pattern under the parameterisation examined.

\subsection{Flat-landscape control}

We next test whether the observed transition-network growth depends on the existence of dynamically meaningful basin structure.

\begin{figure}[H]

    \centering

    \includegraphics[width=\textwidth]{\detokenize{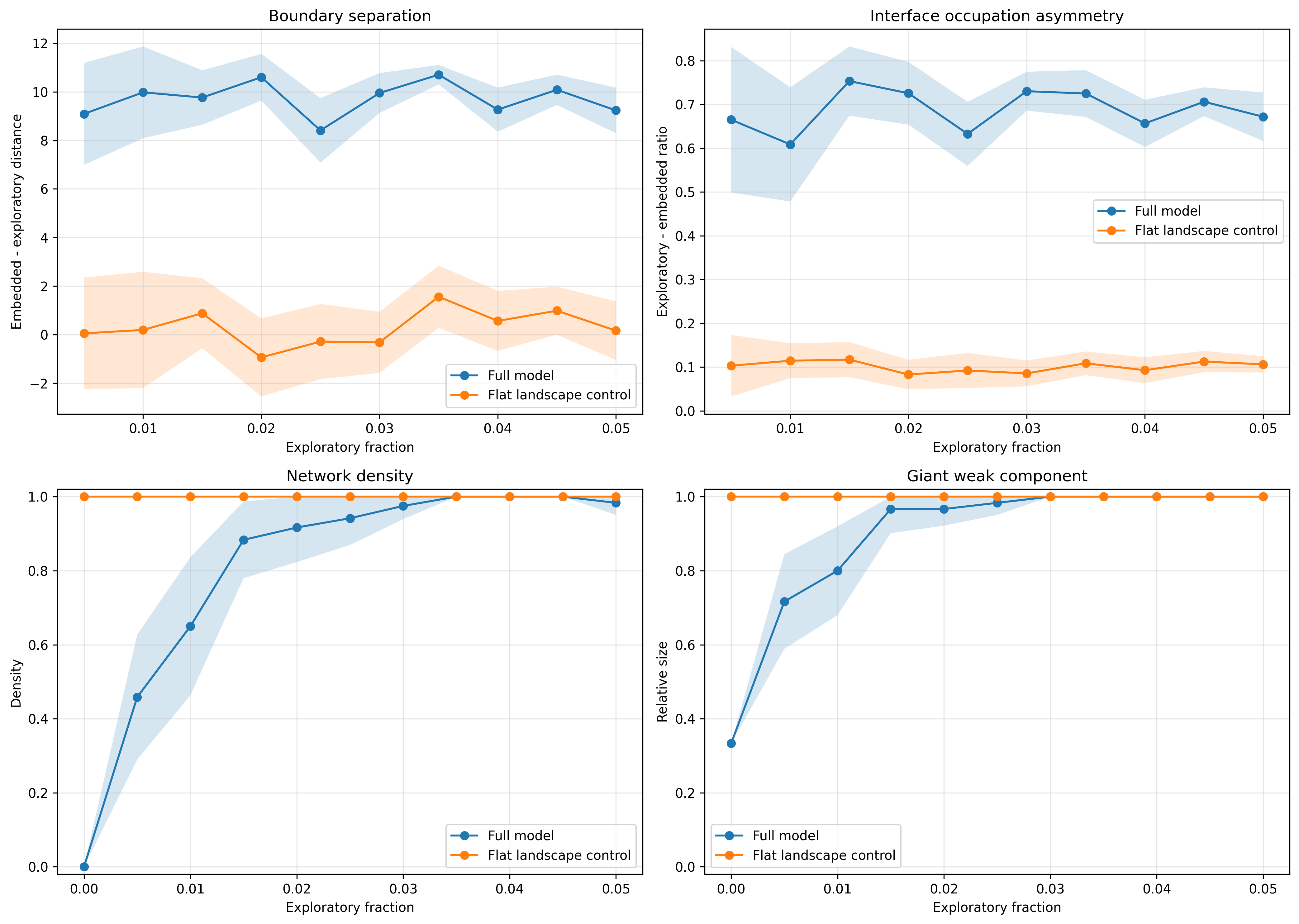}}

    \caption{Flat-landscape control. Comparison between the full structured model and a control in which attractor-following gradients, interface-sensitive selection, and novelty-dependent selection are disabled, while the same nearest-attractor partition is retained for state assignment and transition detection. Panels compare boundary-distance separation, interface-occupation asymmetry, transition-network density, and the relative size of the largest weakly connected component. Shaded regions are approximate 95\% confidence intervals across independent runs; bands for bounded network measures are clipped to \([0,1]\).}

    \label{fig:flat_control}

\end{figure}

As shown in Figure~\ref{fig:flat_control}, removing attractor-following gradients eliminates the dynamical confinement that distinguishes basin interiors from interfaces. Agents can then move across the operational nearest-attractor partitions without crossing dynamically separated basins, and transition-network connectivity is already saturated at \(p=0\) in the reported control.

The control therefore does not show that spatial structure is required for connectivity in a generic sense. Rather, it shows that the scientific problem addressed here is connectivity generated \emph{despite} local confinement. In the structured landscape, global integration emerges while embedded agents remain spatially localized and exploratory agents remain disproportionately exposed to basin interfaces. In the flat control, by contrast, connectivity reflects unconstrained spatial mixing rather than boundary-mediated integration of structured regions.

\section{Discussion}

\subsection{A minimal boundary-mediated mechanism}

Our results identify a minimal mechanism through which heterogeneous mobility generates global connectivity in a structured landscape. The mechanism does not require optimisation, learning, or goal-directed behaviour, but arises from the interaction between stochastic movement and environmental structure.

Exploratory and embedded agents differ only in their movement dynamics. In a structured landscape, this heterogeneity produces different patterns of spatial sampling, while transitions between attractor states remain strongly localised near basin interfaces. Boundary localisation itself is geometric and occurs across all exploratory mechanisms. Interface sensitivity changes the frequency with which agents occupy these transition-enabling regions and raises the crossing rate by approximately four- to fivefold per exploratory movement. Repeated boundary crossings then accumulate into an expanding transition network, linking local movement to global connectivity.

The mechanism can therefore be summarised as

\[
\begin{aligned}
&\mathrm{heterogeneous\ mobility}
\;\times\;
\mathrm{structured\ landscape}
\\
&\qquad\longrightarrow\;
\mathrm{interface\ exposure}
\;\longrightarrow\;
\mathrm{boundary\ crossings}
\\
&\qquad\longrightarrow\;
\mathrm{global\ connectivity}.
\end{aligned}
\]

This generative picture is consistent with broader accounts of emergence in which system-level organisation arises from local dynamics operating under constraints \cite{holland1998,kauffman1993,scheffer2009,prokopenko2009,epstein1999}. Here, interfaces provide the spatial loci through which local stochastic dynamics become systemically consequential.

The robustness analyses support this interpretation across changes in attractor number, landscape scaling, simulation duration, and population size (Supplementary Material). Connectivity growth persists across increasing numbers of attractors when basin separation is preserved. Under fixed-domain scaling, however, increasing \(K\) progressively crowds the landscape, reduces geometric basin separation, and increases spontaneous transition activity. The relevant constraint is therefore not attractor number per se, but the geometry and separation of the basins through which transitions occur.

Finite-time analyses further show that incomplete connectivity in larger state spaces should not be interpreted automatically as a failure of the mechanism. For \(K=50\), increasing the observation horizon produces progressively greater global integration, indicating that larger transition networks require additional time for local crossing events to accumulate into system-spanning connectivity.

\subsection{From local crossings to global connectivity}

Global connectivity emerges through the accumulation of local transition events rather than through homogeneous link formation. As the exploratory fraction increases, boundary crossings generate new links between attractor regions, progressively integrating the transition network. The movement-normalized analysis shows that this is not solely a consequence of providing more exploratory agents or more total steps: interface-sensitive agents cross basin boundaries much more frequently per movement than unbiased random walkers.

The resulting connectivity growth is percolation-like in the sense that a system-spanning connected component emerges from accumulated local events \cite{newman2018,barabasi2016,boccaletti2006,stauffer2003}. Unlike classical percolation on a predefined network, however, the graph here is generated endogenously by agent trajectories, and its links arise from spatial transitions between attractor regions.

Population-scaling experiments further suggest that the absolute number of exploratory agents is an important control variable in addition to their relative fraction. A fraction sweep shows population-size dependence at fixed \(p\), while a separate experiment sets common target expected explorer counts through \(p=N_e^{\mathrm{target}}/N\). Connectivity curves for different population sizes align more closely under this common-target design (Supplementary Material). This scaling is consistent with a mechanism in which the number of agents capable of generating cross-basin transitions influences the rate at which transition pathways accumulate. A separate Bernoulli-assignment diagnostic confirms that the realised exploratory counts track their expected values across the parameter range.

\subsection{Beyond strategic exploration}

A central implication of the results is that global reconfiguration does not require agents to actively seek novelty or optimise their trajectories. The persistence of connectivity growth under pure random-walk exploration shows that directed search and adaptive strategies are not necessary for the core mechanism.

This contrasts with models in which exploration and innovation are associated with learning, heuristics, or strategically guided search \cite{march1991,levinthal1997,gavetti2000}. In the present model, exploratory behaviour can instead arise operationally from differences in stochastic mobility interacting with environmental constraints.

Interface-sensitive candidate selection localises exploratory activity more strongly at basin interfaces and produces the dominant rate increase. The full model exhibits an additional increase of approximately 3.5--11.5 transitions per \(10^3\) movements at six of the ten positive exploratory fractions under pointwise paired bootstrap intervals. This secondary effect is small relative to the interface-only versus baseline contrast, is not uniform across \(p\), and does not translate into a consistent increase in distinct-edge production or unique-edge yield. Novelty is therefore not required for the emergence of connectivity and is not supported as a separate topological mechanism, although it can modestly alter crossing frequency under some conditions. This is consistent with generative approaches in which macro-level organisation arises from comparatively simple local processes \cite{holland1998,prokopenko2009,epstein1999}.

\subsection{The role of landscape structure}

The flat-landscape control reinforces the distinction between generic spatial mixing and boundary-mediated connectivity in a structured landscape. When attractor-following gradients are removed, agents can still cross the operational partitions used for state assignment, but these crossings no longer represent transitions between dynamically separated basins.

Connectivity therefore remains measurable in the flat landscape, but its interpretation changes: it reflects unconstrained spatial mixing rather than the progressive integration of structured regions. The relevant distinction is consequently not whether connectivity exists, but how it is generated.

Environmental structure thus plays a dual role. It constrains movement while simultaneously creating the interfaces through which movement can generate meaningful state transitions. This role of constraints is consistent with broader accounts of self-organisation and complex adaptive systems \cite{holland1998,kauffman1993,scheffer2009,prokopenko2009}.

\subsection{Implications for complex systems}

The model is deliberately abstract and is not calibrated to reproduce a particular empirical system. Its constituent processes are nevertheless not empirically arbitrary. Human-mobility studies reveal the coexistence of recurrently localised and more exploratory trajectories, including an empirically identified returner--explorer dichotomy with distinct consequences for spreading and social interaction \cite{song2010,pappalardo2015,alessandretti2018}. These observations provide a behavioural counterpart for the model's mobility heterogeneity, although the empirical categories are inferred from observed trajectories whereas agent types are assigned exogenously here.

Scientific-career data provide a second correspondence. Changes in research interests and disciplinary classifications generate measurable flows between cognitive domains \cite{jia2017,dedomenico2016,aleta2019,eykens2023}. In particular, Eykens et al. construct a network from researchers' disciplinary switches and report that such moves are not confined to the most cognitively similar disciplines \cite{eykens2023}. This empirical construction closely parallels the transition-network representation used here: differentiated disciplines provide domains, career switches provide boundary-crossing events, and their aggregation provides a network of inter-domain movement.

The final correspondence concerns global topology. In a society-scale mobile-communication network, weak inter-community ties were found to be disproportionately important for maintaining global connectivity: their ordered removal fragmented the network through a phase transition \cite{onnela2007}. In urban mobility, Zhang et al. used empirical origin--destination flows to identify the smallest rank cutoff yielding a strongly connected backbone and found a stable percolation scale across cities and observation periods \cite{zhang2026}. These studies support the structural relevance of sparse cross-domain connections and percolation-based descriptions, but they do not test the generative process implemented in the present model.

Taken together, this literature provides empirical grounding for the model ingredients and qualitative support for its structural logic. It does not constitute quantitative validation of the simulated threshold, parameter values, or causal mechanism. The contribution of the present model is instead to show generative sufficiency: under controlled assumptions, heterogeneous mobility interacting with a structured landscape is sufficient to convert local state changes into an increasingly connected transition network.

In social and organisational settings, mobile individuals can connect otherwise separated domains, with potential consequences for knowledge diffusion and innovation. This interpretation is related to work on exploration--exploitation, search across complex landscapes, and bridging between weakly connected groups \cite{march1991,levinthal1997,gavetti2000,granovetter1973,burt2004,uzzi2005,onnela2007}.

Analogous dynamics may arise in ecological or evolutionary systems when movement connects niches, patches, or regions of an adaptive landscape \cite{kauffman1993,huang2012}. These parallels should be interpreted as potential applications of the mechanism rather than direct empirical predictions. The general result is that a relatively small mobile subpopulation can disproportionately affect connectivity when its trajectories cross boundaries separating otherwise structured regions.

\subsection{Limitations and future directions}

The deliberate minimality of the model also defines its limitations. Agents do not adapt their behaviour over time, the attractor landscape is externally specified and static, and interactions between agents are mediated through the shared spatial environment rather than explicit communication, competition, or cooperation. Extending the framework to adaptive agents, co-evolving landscapes, or explicit interaction networks would test how the minimal mechanism operates under richer dynamics.

A second limitation concerns the interpretation of the connectivity transition. The present results support a percolation-like description but do not establish a formal critical phenomenon. Larger-scale finite-size analyses and analytical approximations would be required to determine whether the observed behaviour belongs to an established percolation class or instead represents a distinct dynamical connectivity process \cite{newman2018,barabasi2016,boccaletti2006,stauffer2003}.

A third limitation is the rapid topological saturation of the main \(K=3\) transition graph, which contains at most six directed edges. The movement-normalized distinct-edge production and unique-edge yield therefore decline as repeated transitions accumulate, even when connectivity is already complete. The \(K\)-robustness analysis shows that connectivity growth persists in larger landscapes, but applying the same event-normalized edge-discovery analysis to larger \(K\) would be required to determine whether exploratory mechanisms differ in route diversity at a common transition budget. In addition, the pointwise paired bootstrap intervals used for the mechanism contrasts are descriptive and unadjusted for multiple comparisons; the large, uniform interface-only versus baseline rate effect is robust, whereas isolated novelty effects should be treated as exploratory.

The empirical correspondences above also suggest direct tests. If the proposed mechanism operates in observed systems, cross-domain transitions should be disproportionately generated by the more mobile part of the population; aggregate connectivity should be organised more consistently by the realised number of such individuals than by their nominal population fraction alone; and interface-sensitive movement should accelerate, but not be necessary for, the formation of cross-domain pathways. Testing these predictions would require longitudinal trajectories, an independently specified domain partition, and time-resolved transition networks. Such tests would move beyond the qualitative empirical grounding established here toward direct validation or falsification of the proposed mechanism.

\subsection{Concluding perspective}

Global connectivity in this model arises from a simple sequence: heterogeneous movement brings a subset of agents into contact with basin interfaces, interface exposure increases boundary-crossing frequency, local crossings generate transitions between attractor regions, and their accumulation progressively integrates the transition network.

The central result is therefore not that exploratory agents require sophisticated strategies, but that simple stochastic mobility can become systemically effective when embedded in an appropriately structured environment. Interface sensitivity is a sufficient local rule that accelerates this process by increasing crossings per unit movement, while novelty is not required. Interfaces provide the link between local dynamics and global organisation, converting repeated movement into system-wide connectivity.

\section*{Supplementary Material}

Supplementary Methods, robustness and scaling analyses, diagnostic controls,

and illustrative transition-network snapshots are provided in the accompanying

Supplementary Material.

\section*{Acknowledgements}

A preprint version of this manuscript has previously been published on arXiv

\cite{nelli2026arxiv}.

\section*{Use of Artificial Intelligence}

Generative AI tools were used after preparation of the initial

manuscript, model, implementation, simulations and analyses to assist

with English-language editing, LaTeX preparation, reviewer-style

feedback, and diagnostic consistency checks among the manuscript,

mathematical notation, reported figures and Python implementation.

AI tools also suggested possible improvements in the presentation of

robustness analyses. All suggestions were critically evaluated by the

author. Any resulting code modifications and additional analyses were

independently implemented, executed and verified by the author. The

research question, original model design, simulation data, scientific

interpretation and final conclusions were determined by the author,

who reviewed the complete manuscript and takes full responsibility

for the work.

\section*{Code and Data Availability}

All simulations and analyses presented in this study were implemented in Python.

The complete source code, experimental pipeline, configuration files, data

processing scripts, and figure-generation scripts are publicly available.

The source code is available at

\href{https://github.com/meccanismocomplesso/boundary-connectivity-abm}{GitHub}.

A curated release of the model is available through the

\href{https://www.comses.net/codebases/20606999-844a-40ca-b423-1a5404453d83/}{CoMSES Net Computational Model Library}.

A citable archived version of the code is available on Zenodo at

\href{https://doi.org/10.5281/zenodo.23864812}{DOI: 10.5281/zenodo.23864812}.

An archived version of the manuscript is available on Zenodo at

\href{https://doi.org/10.5281/zenodo.20145381}{DOI: 10.5281/zenodo.20145381}.

A public preprint version of the manuscript is available on arXiv at

\href{https://arxiv.org/abs/2606.07344}{arXiv:2606.07344}.

The repository contains the implementation required to reproduce the simulations,

numerical analyses, and figures reported in the manuscript. Random seeds and

model parameters are explicitly controlled and recorded to support independent

replication.

\section*{Competing Interests}

The author declares no competing interests.

\section*{Funding}

This research received no external funding.

\section*{Author Contributions}

Fabio Nelli: Conceptualization, Methodology, Software, Formal Analysis,

Investigation, Validation, Visualization, Data Curation, Writing---Original Draft,

Writing---Review and Editing.

\end{document}